\documentclass[aps,prl,epsfig,twocolumn,showpacs]{revtex4}
\usepackage{amsmath}
\usepackage{graphicx}

\begin{document}
\title{ On the effective one-component description of highly asymmetric
hard-sphere binary fluid mixtures}

\author{C. F. Tejero}
 \email[]{cftejero@fis.ucm.es}
\thanks{ permanent address: Facultad de Ciencias F\'{\i}sicas, Universidad Complutense de
Madrid, E-28040, Madrid, Spain}
\author{M. L\'opez de Haro}
\email[]{malopez@servidor.unam.mx}
\homepage[]{http://miquiztli.cie.unam.mx/xml/tc/ft/mlh/}
 \affiliation{Centro de Investigaci\'{o}n en Energ\'{\i}a,
U.N.A.M., Temixco, Morelos 62580 (M{\'e}xico)}

\begin{abstract}
\indent {The phase diagram of  a binary fluid mixture of highly
asymmetric additive  hard spheres is investigated. Demixing is
 analyzed from  the exact  low-density expansions of the
thermodynamic properties of the mixture and compared with the
fluid-fluid separation based on the effective one-component
description. Differences in the results obtained from both
approaches, which have been claimed to be equivalent, are pointed
out and their possible origin is discussed. It is argued that to
deal with these differences new theoretical approximations should
be devised.}
\end{abstract}
\pacs{05.70.Fh, 64.10.+h, 64.70.Ja, 64.75.+g}

\maketitle

It is well known that, due to the widely different time and length
scales involved, it is a very difficult task to describe a soft
matter complex fluid composed of mesoscopic (colloidal) particles
and microscopic (small) particles, i.e. a colloidal suspension.
Therefore people have struggled for years trying to devise
strategies that may render a tractable description. On the
theoretical side, a rather successful and often employed approach
in the colloids literature is that of coarse graining. In this
technique the idea is to integrate out the irrelevant degrees of
freedom (the small particles) to end up with a one-component
system of colloidal particles described by effective interactions
\cite{Likos}. One of the most celebrated effective interactions
results from the depletion mechanism first described by Asakura
and Oosawa \cite{Asakura}. Suppose that two colloidal hard spheres
are immersed in an ideal fluid of small hard spheres. When the
separation of the surfaces of two large spheres is less than the
diameter of the small ones, the depletion of the latter from the
gap between the colloids leads to an unbalanced osmotic pressure
which, in turn, results in an effective attraction between the two
large spheres. Although the Asakura-Oosawa depletion pair
potential is concerned only with the limit of extreme dilution,
the theory of the depletion mechanism has recently been reexamined
by also including the interactions between the small spheres
\cite{Gotzelmann,Roth}. In several of the recent studies the
depletion mechanism has been invoked as the possible driving force
for demixing, i.e. a phase transition whereby a binary fluid
mixture of hard spheres separates into two fluid phases of
different composition.

The aim of this paper is to shed some more light on the usefulness
or limitations of the depletion pair potential picture, i.e. the
consideration of an effective one-component fluid of large spheres
with effective pair interactions for describing the phase behavior
of a binary fluid mixture of highly asymmetric additive hard
spheres \cite{DRE99}. Such an issue has recently been addressed by
simulations \cite{BK98,B99,LW99,MH01} but the analysis
unfortunately provides no definitive answer due to limitations in
the presently available simulation algorithms.

Recently \cite{LdHT04} we have analyzed  the demixing transition
of a binary fluid mixture of $N= N_1 + N_2$ additive hard spheres
of diameters $\sigma_1$ and $\sigma_2$ ($\sigma_1>\sigma_2$). For
a fixed diameter ratio $\gamma= \sigma_2/\sigma_1$, the
thermodynamic properties of the mixture were described in terms of
the number density $\rho=N/V$, with $V$ the volume, the partial
number fraction of the big spheres $x=N_1/N$, and the total
packing fraction $\eta= \eta_1 + \eta_2$, where
$\eta_1=\pi\rho\sigma_1^3 x/6$  and
$\eta_2=\pi\rho\sigma_1^3(1-x)\gamma^3/6$ are the packing
fractions of the large and the small spheres.  Demixing was
studied by starting from the exact low-density expansions of the
thermodynamic properties, using as input analytical expressions
\cite{Kihara1,Kihara2} and numerical computations
\cite{Saija,Enciso1,Enciso2,Wheatley,Yu,Vlasov} for the virial
coefficients so far reported in the literature (up to the sixth).
For all the size asymmetries considered we found that, by
successively incorporating in the density expansions a new exact
virial coefficient, the critical consolute point moves to higher
values of the pressure and of the total packing fraction as shown
in Fig. $1$. It is also seen in this figure that the rate of
convergence is slow and so definitive conclusions on the location
of the critical consolute point cannot yet be reached with the
available data on virial coefficients.  We have also compared
these findings with the fluid-fluid separation resulting from
different empirical proposals for the equation of state
\cite{Coussaert,Hamad,Santos2}, that attempt to reproduce the
virial behavior and/or to comply with consistency conditions of
the contact values of the radial distribution functions. We have
realized that for historical reasons all these ``rescaled''
equations of state yield the exact second and third virial
coefficients but inherit the singularity of the Percus-Yevick
theory. We have shown that this apparently innocuous singularity
arbitrarily introduces, as compared to the exact low-density
expansion of the pressure, additional dependencies on $x$ through
the  fourth, fifth, and so on virial coefficients. Such
dependencies make the location of the critical consolute point to
vary enormously from one equation of state to another, so that the
predictions stemming from each of them are indeed unreliable.

\begin{figure}[tbp]
\includegraphics[width=.80 \columnwidth]{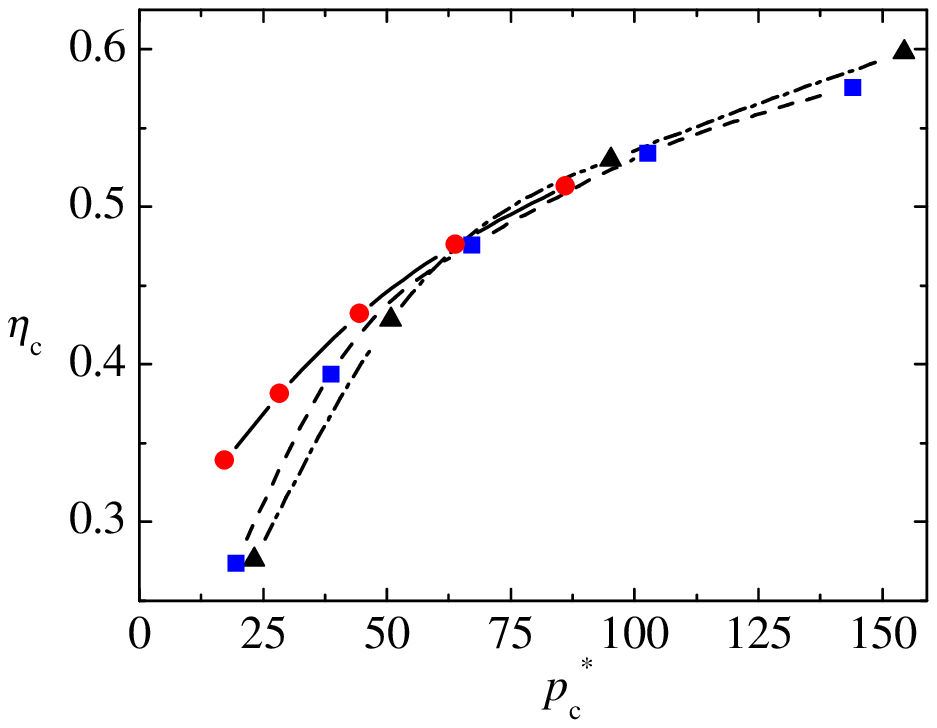}
\caption{Critical total packing fraction $\eta_c$ as a function of
the critical reduced pressure $p_{c}^{*}$, where $p^* \equiv \beta
p \sigma_{1}^3$, $\beta = 1/k_B T$, with $k_B$ Boltzmann's
constant, and $T$ the absolute temperature, as obtained from the
exact low-density expansions for three values of the diameter
ratio $\gamma$. Successive points (from left to right) indicate
the addition of  a new exact virial coefficient, starting with the
second.  Lines have been drawn to guide the eye. Dash-dotted line
and triangles: $\gamma=0.05$; dashed line and squares:
$\gamma=0.1$; solid line and circles: $\gamma=0.2$. \label{fig1}}
\end{figure}

The demixing transition of a binary mixture of additive hard
spheres may be analyzed using different thermodynamic planes. In
order to compare to well documented results in the effective
one-component fluid, we first consider the ($\eta_1-\eta_2$)
plane. In Fig. $2$ we present the binodals resulting from the
exact low-density expansions up to the sixth virial coefficient
for two size asymmetries, namely $\gamma=0.2$ and $\gamma=0.1$,
and up to the fifth virial coefficient for $\gamma= 0.05$. Note
that the density region spanned by each binodal cannot go beyond
the point where the total packing fraction occupied by the spheres
reaches for the dense phase the highest possible value. Since
these values are not known, we have drawn all the curves up to the
limiting (unphysical) value $\eta=1$. A comparison between these
binodals and parallel ones derived using the depletion mechanism
\cite{DRE99} indicates the following. For $\gamma=0.2$, the
low-density expansion predicts demixing, whereas there is no
fluid-fluid separation in the effective one-component fluid. The
qualitative behavior of the binodals for $\gamma= 0.1$ and
$\gamma= 0.05$ is, on the other hand, rather similar in both
approaches. However, there are two main differences:  1) the
values of $\eta_2$ for the mixture are much lower than the
corresponding ones in the effective one-component fluid; 2) the
critical consolute points already lie outside the range where the
binodals are predicted in the effective one-component fluid.

\begin{figure}[tbp]
\includegraphics[width=.80 \columnwidth]{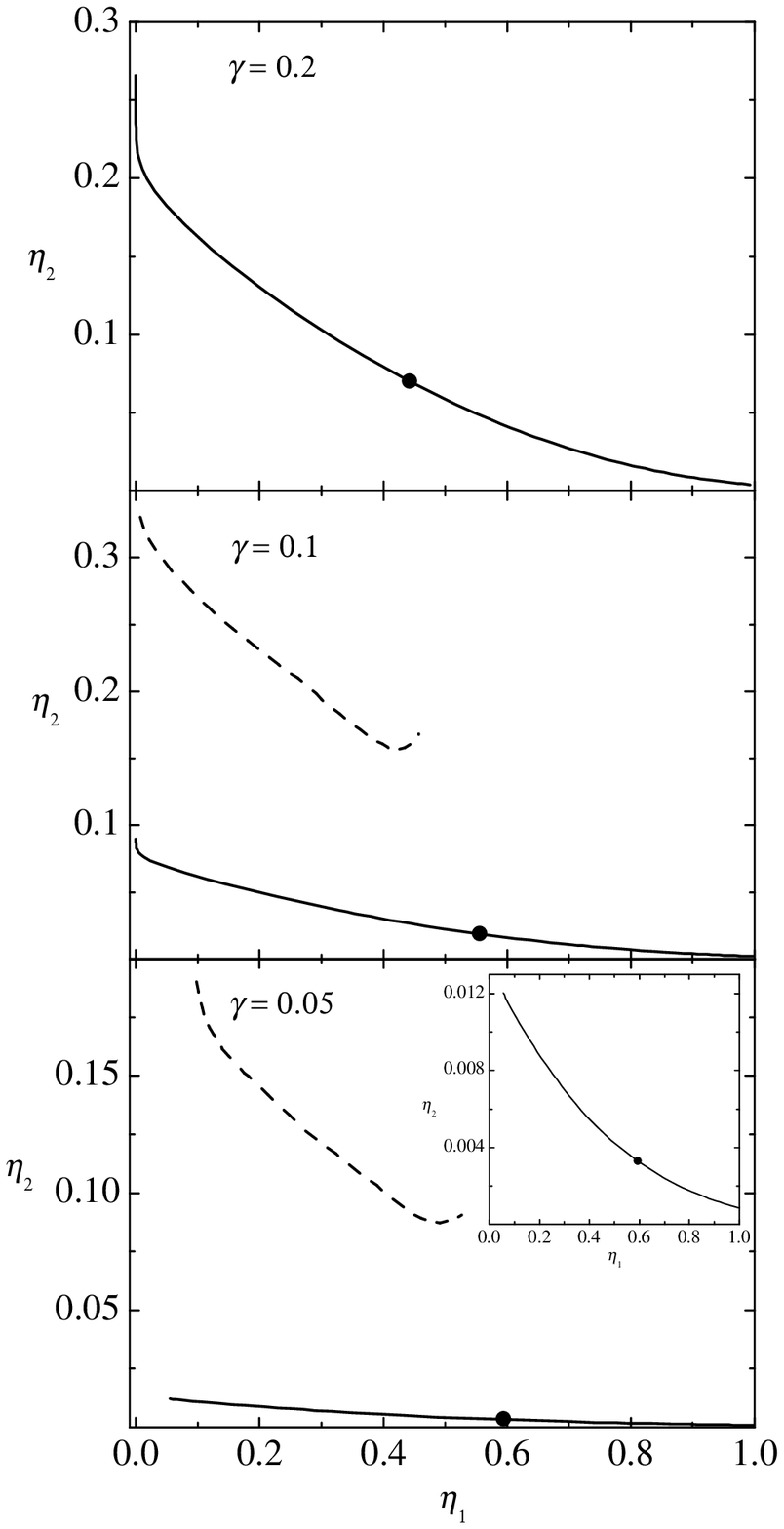}
\caption{ Binodals of  asymmetric binary fluid mixtures of
additive hard spheres in the ($\eta_1$-$\eta_2$) plane as obtained
from the exact low-density expansions (continuous lines) for the
same values of the diameter ratio as considered in Fig. $1$.
Filled circles indicate the critical consolute points. The dashed
lines are the results of Ref. \protect\onlinecite{DRE99}. The
inset in the lower pannel is the curve for $\gamma= 0.05$ in a
magnified scale. \label{fig2}}
\end{figure}

Figure $3$ is yet another presentation of the  binodals,
 this time in the ($\eta_1- p$) plane, with $p$ the
pressure.  These binodals follow from the low-density expansions
by keeping the same number of exact virial coefficients as in Fig.
2. Since no values of the pressure for the  coexisting phases have
been reported in the effective one-component fluid in
\cite{DRE99}, a direct comparison between both approaches is not
possible in this case. Nevertheless, the following comments are in
order. It is known that a one-component fluid with pairwise
interactions consisting of a repulsive part and a short-range
attraction  can only produce in the ($\eta_1- p$) plane a van der
Waals-like fluid-fluid separation ending in an  upper critical
point, the concavity of the  binodals pointing downwards.
Moreover, it is well established that, by strongly reducing the
range of the attractions with respect to that of the repulsions,
the usual phase diagram with two stable transitions, namely
fluid-fluid and fluid-solid, changes and the fluid-fluid
transition becomes metastable with respect to the fluid-solid
transition \cite{Gast}. By a further reduction of the range of the
attractions, a stable isostructural solid-solid transition
develops at high densities in the phase diagram
\cite{Frenkel,Tejero}. The phase behavior found in Ref.
\cite{DRE99} follows indeed this trend. Note that although the
zero-body and one-body terms (the volume terms) in the effective
one-component description play no role in determining the phase
equilibria, they are however responsible for inverting the
concavity of the binodals from downwards to upwards, yielding the
same concavity as the one found in the true binary mixture. The
importance of the volume terms in the determination of the
complete phase diagram should therefore not be ignored
\cite{Likos}. On the other hand, the resulting phase diagrams
found in Ref. \cite{DRE99} also indicate that, for the depletion
pair potential used in this reference, the small repulsive barrier
that follows the attractive potential close to the surface of the
large sphere does not  qualitatively modify the phase diagram that
would result by only considering a purely short-range attractive
potential. The predictions for the phase diagram in the effective
one-component fluid  \cite{DRE99} for $\gamma=0.1$ (a metastable
fluid-fluid transition) and for $\gamma= 0.05$ (a metastable
fluid-fluid transition and a stable solid-solid transition) are
indeed at grips with the foregoing remarks.

\begin{figure}[tbp]
\includegraphics[width=.80 \columnwidth]{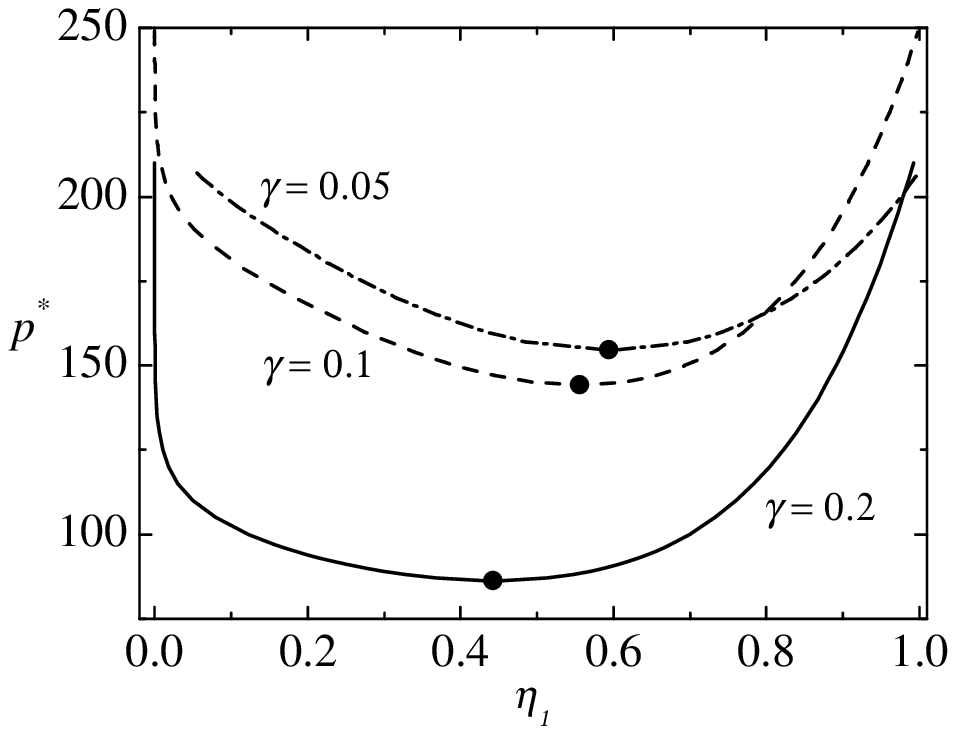}
\caption{ Binodals  of  asymmetric binary fluid mixtures of
additive hard spheres in the ($\eta_1$-$p^*$) plane as obtained
from the exact low-density expansions for the same values of the
diameter ratio as considered in Figs. \protect\ref{fig1} and
\protect\ref{fig2}. Dash-dotted line: $\gamma=0.05$; dashed line:
$\gamma=0.1$; solid line: $\gamma=0.2$. Filled circles indicate
the critical consolute points. \label{fig3}}
\end{figure}

\begin{figure}[tbp]
\includegraphics[width=.90 \columnwidth]{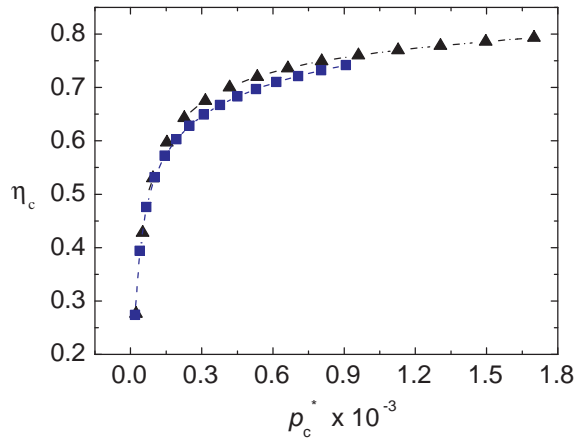}
\caption{Critical total packing fraction $\eta_c$ as a function of
the critical reduced pressure $p_{c}^{*}$ as obtained from the
truncation of the  virial expansion of the BMCSL equation of state
for two of the values of the diameter ratio $\gamma$ considered in
Fig. \protect\ref {fig1}. Successive points (from left to right)
indicate that an additional (approximate) virial coefficient has
been retained in the truncation, starting with the second. Lines
have been drawn to guide the eye. Dash-dotted line and triangles:
$\gamma=0.05$; dashed line and squares: $\gamma=0.1$.
\label{fig4}}
\end{figure}

It should be clear that due to the fact that only a few exact
virial coefficients are available and that the radius of
convergence of the virial series in the case of mixtures is not
known, definite conclusions derived from the low-density
expansions cannot be drawn. Nevertheless, the observed trends lead
to the following considerations. As shown in Fig. 1, as one
increases the number of terms in the low-density expansions, both
the critical pressure and the critical total packing fraction
grow. It is not clear whether this feature will change by
including more exact virial coefficients (when available) in the
density expansions, but all the evidence so far makes it unlikely.
In any event, provided the trend obtained with the low-density
expansions is a true feature, two possible scenarios arise.  One
may envisage the convergence of the critical constants to a
thermodynamic state with either a physical or an unphysical
critical total packing fraction. Indeed, the terms physical and
unphysical depend of the resulting solid phase at high densities.
For example, if the freezing transition corresponds to the partial
freezing of the large spheres, i.e. a face-centered cubic lattice
formed by the large spheres while the small spheres remain
disordered \cite{Coussaert}, the close packing fraction of the
large spheres $\eta_1^{\rm cp}\simeq 0.74$ defines the boundary
between physical and unphysical. In the first scenario the
demixing transition would be thermodynamically metastable with
respect to the freezing transition as found in previous studies,
while in the second no demixing would be present at all. At this
stage we have no way to decide which of the two scenarios is
correct since, by incorporating up to the fifth (sixth) virial
coefficient in the exact low-density expansions for $\gamma=0.05$
($0.1$),
 the critical packing fraction of the big
spheres $\eta_{1}^{c}$ is $\eta_{1}^{c}= 0.595 (0.557)$.
Concerning this point, a related analysis carried out with the
Boubl\'{\i}k-Mansoori-Carnahan- Starling-Leland (BMCSL) equation
of state \cite{Boublik} (which predicts no deximing) is
illustrative. This is shown in Fig. \ref{fig4}. The striking
feature to be noted is the excellent agreement of the results
obtained with the exact low-density expansion ({\em c.f.} Fig.
\ref{fig1}) and those obtained with the truncated BMCSL equation
of state. By including additional virial coefficients of the BCMSL
equation of state it is seen that $\eta_c\simeq \eta_{1}^{c}$
exceeds the close-packing fraction of the large spheres.

In order to understand the differences between the phase diagrams
presented in Fig. 2 using the exact low-density expansions and the
effective one-component description, the following considerations
should be pointed out. For $\gamma= 0.2$ both approaches would be
equivalent whenever the critical packing fraction of the big
spheres resulting from the low-density expansions converges to an
unphysical value. As indicated above, for $\gamma= 0.05$ and
$\gamma= 0.1$ the critical consolute points obtained with a
reduced number of exact virial coefficients already lie outside
the range where the effective one-component description locates
the binodals. In these cases the effective one-component
description would only provide a qualitative approximation of the
true binary mixture if $\eta_{1}^{c}$ converges to a physical
value. Otherwise (convergence to an unphysical value) the
predictions of both approaches would be completely at odds with
each other in this basic issue.  All these possibilities can
indeed only be tested by including more exact virial coefficients
in the density expansions. As a possible source for the
differences found between the exact density expansions and the
effective one-component description we suggest that it may be due
to the fact that, in order to determine in the latter the phase
diagram and to perform the direct simulations \cite{DRE99}, the
Carnahan-Starling equation of state is used for describing both
the big and the small hard spheres. Note that the
Percus-Yevick-type singularity of the Carnahan-Starling equation
of state may induce for each component different (through $\eta_1$
and $\eta_2$) spurious $x$-dependences similar to those discussed
in Ref. \onlinecite{LdHT04} in connection with equations of state
derived from ``rescaled'' virial expansions. These dependencies
would in turn lead to unreliable predictions on the location of
the demixing transition.

Up to here and due to the involved approximations, none of the
arguments and results that we have discussed lead to a neat
conclusion concerning the claimed equivalence of the effective
one-component description vis a vis the true binary mixture
description of highly asymmetric hard spheres. Therefore,  new
theoretical approaches to deal with this  problem should be
sought. In this respect, systematic and controlled approximations
in the fluid regime, may profitably be undertaken. Of course these
are presently limited by the available virial coefficients. Also,
the role of reliable simulations capable of exploring the region
of interest can hardly be overemphasized. Irrespective of the
technical difficulties that no doubt are involved both in the
simulations and in the computation of additional virial
coefficients of highly asymmetric binary fluid mixtures of
additive hard spheres, we hope that the present paper may also
serve to stimulate efforts in this direction.

C.F.T. acknowledges financial support from  the Mi- nisterio de
Ciencia y Tecnolog\'{\i}a (Spain) Ref: BFM2001-1017-C03-03. This
work has also benefitted from the Collaboration Agreement between
Universidad Nacional Aut\'{o}noma de M\'{e}xico and Universidad
Complutense de Madrid.




\end{document}